\documentstyle[aps,preprint]{revtex}

\begin{document}
\draft

\title{Universality Classes,
Statistical Exclusion Principle and  Properties of Interacting Fermions}

\author{Krzysztof Byczuk$^{a}$ and Jozef Spa\l ek$^{b,a}$
\footnote{E-mail: byczuk@fuw.edu.pl, ufspalek@if.uj.edu.pl}}

\address{(a) Institute of Theoretical Physics, Warsaw University,
ul. Ho\.za 69, 00-681 Warszawa, Poland \\
(b) Institute of Physics, Jagellonian University, ul. Reymonta 4, 30-059
Krak\'ow, Poland }

%\date{today}
\maketitle

\begin{abstract}
We point to the possibility of existence of the statistical-spin-liquid
state as the state
 which differs from either Fermi or Luttinger liquid states.
In the statistical spin liquid the double occupancies are excluded from
the physical space.
Each of the above
three cases (Fermi, Luttinger and spin liquids)
 represents an universality class for the interacting
many-particle fermion systems.
The properties of the spin liquid such as the chemical potential, the entropy,
and the magnetization curve, as well as the quasiparticle
structure are briefly discussed.
\end{abstract}

%\newpage

%\begin{narrowtext}

In the theory of correlated systems the principal problem is to transform
the microscopic model of interacting particles into an effective approach
with interaction among quasiparticles and/or collective excitations.
Haldane \cite{hal1}
in his seminal paper provided an unified framework of the effective
theory of interacting fermions in a normal (metallic) state.
The Fermi surface, whose existence is a basic postulate of the approach,
is considered as a d-1 dimensional collection of points, where the
momentum distribution  function $n_{\bf k}$ has singularities.
Obviously, this might be either a step discontinuity (as in the
Fermi liquid case), or any weaker singularity (as in the
case of Luttinger liquid).
The Fermi surface in both cases must obey the Luttinger theorem which is
taken as an additional postulate.
Starting from these two postulates one can describe the low-energy
long-wave-length excitations of the system solely in terms of the fluctuating
Fermi surface.

Explicitly, one introduces the local Fermi wave vector ${\bf k}_{F {\bf Q}
\sigma} ({\bf x})$ related to the undistorted Fermi wave vector
${\bf k}_{F {\bf Q} \sigma}^{0} ({\bf x})$ by
\begin{equation}
{\bf k}_{F {\bf Q} \sigma} ({\bf x}) =
{\bf k}_{F {\bf Q} \sigma}^{0} ({\bf x}) +
\delta {\bf k}_{F {\bf Q} \sigma} ({\bf x}),
\end{equation}
where vector ${\bf Q}$ with d-1 components, where d is the space dimension,
 describes a
coarse-grained point of the Fermi surface, $\sigma=\pm 1$ is the spin quantum
number, and
${\bf x}$ is a position variable.
It turns out \cite{hal1} that only normal to the surface fluctuations
$ \delta {\bf k}_{F {\bf Q} \sigma}^{(\parallel)} ({\bf x})$
give contributions to
the physical quantities such as the local fluctuation in the total
particle density, which is defined as $ \delta n_{\sigma}({\bf x}) =
\sum_{{\bf Q}} \delta n_{{\bf Q} \sigma}({\bf x}) = \sum_{{\bf Q}}
\delta {\bf k}_{F {\bf Q} \sigma} ({\bf x})
\Delta \nu_{{\bf Q}}$, where $\Delta
\nu_{\bf Q}$ characterizes the Fermi surface discontinuity in the ground
state occupation number $n_{\bf k}$ distribution.
Moreover, the normal fluctuations of  the Fermi surface  obey the
commutation  relations of a Kac-Moody algebra \cite{hal1}.
Therefore, one can write down the most general effective Hamiltonian for
interacting fermions in the following bosonized form
\begin{equation}
\Delta H^{eff} = \frac{1}{2} \sum_{\bf q}\sum_{\bf Q_{1}\bf Q_{2}}
\sum_{\sigma  \sigma '} \Gamma ^{\sigma \sigma '}_{\bf Q_{1} \bf Q_{2}}
({\bf q}) \delta n_{{\bf Q}_{1} \sigma}({\bf q})
\delta n_{{\bf Q}_{2} \sigma '}({- \bf q}).
\end{equation}
The operator
$\delta n_{{\bf Q} \sigma}({\bf q})=
\sum_{\bf x} e^{i{\bf qx}} \delta n_{{\bf Q} \sigma}({\bf x})$
is the bosonlike Tomonaga-Luttinger density wave operator
and
$\Gamma ^{\sigma \sigma '}_{\bf Q_{1} \bf Q_{2}}$
is the positive defined matrix of elements given by \cite{hal1}
\begin{equation}
\Gamma ^{\sigma \sigma '}_{\bf Q_{1} \bf Q_{2}}=
\delta_{{\bf Q}_{1}{\bf Q}_{2}} \delta_{\sigma \sigma '} +
\Lambda^{d-1} \frac{f({\bf k}_{F {\bf Q}_{1}\sigma}, {\bf k}_{F {\bf Q}_{2}
\sigma '};{\bf q}) }{ (v_{F {\bf Q}_{1}\sigma} v_{F {\bf Q}_{2} \sigma '}
)^{1/2}}.
\end{equation}
The quantities $v_{F {\bf Q} \sigma}$ are the Fermi velocities on  separate
points of the Fermi surface and the scaling variable
$\Lambda$ is the distance of the
two closest points on it; it is the cut-off parameter in the theory.
An universal behavior of the system depends on how the interaction part
of the effective Hamiltonian (2) behaves in the scaling limit
$\Lambda \rightarrow 0$.
The purpose of this paper is to point out that
in this limit
 a systematic classification
of the interacting fermion liquids into three physically distinct classes
takes place, as well as to specify briefly each of them.

 To address this fundamental
 question we suppose that  the
 interaction part of (2) scales  as
\begin{equation}
\Lambda^{d-1} \frac{f({\bf k}_{F {\bf Q}_{1}\sigma}, {\bf k}_{F {\bf Q}_{2}
\sigma '};{\bf q}) }{ (v_{F {\bf Q}_{1}\sigma} v_{F {\bf Q}_{2} \sigma '}
)^{1/2}} \sim \frac{\Lambda^{d-1}}{\Lambda^{\alpha}},
\end{equation}
where $\alpha$ is the exponent characterizing the type of the singularity.
When $\alpha < d-1$, then the Landau Fermi liquid is stable \cite{hal1}.
For $\alpha = d-1$ the interaction part of the Hamiltonian (2) is finite
and comparable to the kinetic energy when $\Lambda \rightarrow 0$.
Therefore this part is called the relevant variable.
Since this interaction couples two different points of the Fermi
surface, the bosonized
Hamiltonian must be diagonalized via the Bogolyubov
transformation in order to determine the excitation spectrum.
In effect, this leads to the Luttinger liquid type behavior, i.e. the
Luttinger liquid is the stable fixed point for the interacting
electrons in $d$ dimensions \cite{roz}.
However, when $\alpha > d-1$ the interacting part becomes infinite in the
scaling limit $\Lambda \rightarrow 0$ and therefore, the
diagonalization procedure is impossible to perform.

Since the conventional methods are useless in the case when $\alpha > d-1$,
we propose to extend the Pauli exclusion principle in order to
project out from the  Hilbert space the states which lead
to the infinite energy contribution.
More precisely, we take the forward scattering amplitude for
particles with opposite spins as divergent with the exponent $\alpha > d-1$.
Note, that for particles with spin one-half this is the only
nontrivial possibility due to the existence of the
 Pauli exclusion principle, but
for particles with higher spin or additional internal symmetries
(such as orbital degeneracy),
additional
processes may appear in $\Gamma$.
Instead of dealing with the infinite amplitude from the beginning we take
this amplitude to be a finite number $(U_s)$ and put at the
end of calculations $U_s \rightarrow \infty$;
this limit corresponds to the case with a singular scattering
forward amplitude
and, consequently, to the statistical exclusion \cite{s1}
discussed earlier.
Thus the  change of  the total system energy is expressed as follows
\begin{equation}
\delta E = \sum_{{\bf k} \sigma} \epsilon_{{\bf k} \sigma}
< \delta n_{{\bf k} \sigma}> + U_s \sum_{\bf k}
<\delta n_{{\bf k} \uparrow}\delta n_{{\bf k} \downarrow} >
\end{equation}
$$
+\{marginally-relevant-terms \},
$$
where $\epsilon_{{\bf k} \sigma}=\epsilon_{{\bf k}} - \sigma h$ is
the dispersion relation for particles with spin $\sigma$ moving in
an applied magnetic field $h$.
The brackets $<>$ mean taking both  quantum and  thermal
averages.
Within the model defined by the first term in (5), the wave vector ${\bf k}$
is a good quantum number.
Hence, Eq.(5) defines a new effective model
of interacting fermions which is  exactly soluble in the scaling
limit\cite{s1}.
In order to find the exact partition function and the momentum distribution
functions we have to minimize the thermodynamic potential
with respect to both $ < \delta n_{{\bf k} \sigma}>
\equiv \delta n_{{\bf k} \sigma}$ and with
$ <\delta n_{{\bf k} \uparrow}\delta n_{{\bf k} \downarrow} > \equiv
\delta n_{{\bf k} d}$, where
the entropy is defined by\\
$$
\delta S=k_B \sum_{\bf k}[
\sum_{\sigma}(\delta n_{\bf k \sigma}-\delta n_{{\bf k} d})
\ln(\delta n_{\bf k \sigma}-\delta n_{{\bf k} d}) +
\delta n_{{\bf k} d} \ln \delta n_{{\bf k} d} +\\
$$
\begin{equation}
+ (1 - \delta n_{{\bf k} \uparrow} - \delta
n_{{\bf k} \downarrow} + \delta n_{{\bf k} d})
\ln(1 - \delta n_{{\bf k} \uparrow} - \delta n_{{\bf k} \downarrow} +
\delta n_{{\bf k} d})
] .
\end{equation}
The resulting  distribution functions are
\begin{equation}
 \delta n_{{\bf k} \sigma} -
\delta n_{{\bf k} d} =
\frac{ e^{\beta U_s} e^{\beta (\epsilon_{\bf k}-\mu)} \cosh (\beta h)}
{1+e^{\beta U_s}e^{\beta (\epsilon_{\bf k}-\mu)}
[e^{\beta (\epsilon_{\bf k}-\mu)}+2 \cosh(\beta h)]}
\end{equation}
$$
[1 +\sigma \tanh(\beta h)],
$$
and
\begin{equation}
\delta n_{{\bf k} d } =\frac{1}
{1+e^{\beta U_s}e^{\beta (\epsilon_{\bf k}-\mu)}
[e^{\beta (\epsilon_{\bf k}-\mu)}+2 \cosh(\beta h)]}.
\end{equation}
As usually, $\beta$ is the inverse temperature and $\mu$ is the chemical
potential.
Formula (7) describes the probability that the state $|{\bf k}>$ is singly
occupied, whereas Eq.(8)
expresses that of finding a doubly occupied ${\bf k}$
state.

This model has interesting properties as a function of $U_s$.
Namely, at $T=0$ and in the absence of the magnetic field
the distribution function $ \delta n_{{\bf k} \sigma}  $
has two steps for $U<W$, where $W$ is the bandwidth of the single
particles energies $\epsilon_{\bf k}$.
One of the steps is located at the Fermi level $\mu$; below it all
${\bf k}$ states are singly occupied down to the second step located
at energy $(\mu-U_s)$.
For energies below the second step, the doubly occupied ${\bf k}$
configurations
are admissible.

To see the effect of this double-step distribution  function
on the physical properties we calculate the chemical potential $\mu$
as a function of temperature $T$, with the value of $U_s$ as a parameter.
For simplicity we take the constant density of states of the width $W$
and within the energy interval $-W/2 \leq \epsilon \leq W/2$;
the gravity center of the band is chosen at zero energy.
In Fig.1 we present our numerical results for $\mu$.
The case with $U_s=0$ corresponds to the Fermi liquid fixed point.
In that case,
 the chemical potential decreases to $-\infty$ with growing
temperature.
A quite different behavior is observed in the $U_s \rightarrow \infty$
limit.
Then, the chemical potential increases indefinitely.
However, for any intermediate value of $U_s$ the chemical potential
is represented by a retrograde curve, which
starts  decreasing at sufficiently high temperatures.
Thus, even though  for any finite $U_s$ the Fermi surface singularity is
now $\Delta \nu_{\bf Q \sigma} = 1/2$, only the $U_s \rightarrow \infty$
limit leads to a non-Fermi liquid behavior at high temperatures.
This happens because for finite $U_s$ a crossover to the Fermi liquid
behavior takes place whenever $k_BT \geq U_s$.
Also, as displayed in Fig.2, the high-temperature  entropy for
the half filled  band is equal to $2k_B \ln 2$ per particle for
$U_s < \infty$.
As was shown before \cite{s1},
only in the case where the double occupancies are
suppressed entirely,
 the entropy is $k_B\ln 2$ per particle.
The $T\rightarrow \infty$ entropy value $2k_B \ln 2$ per carrier
 characterizes the Fermi liquid
universality class from the statistical point of view, whereas the
value $k_B \ln 2$ is characteristic of a different class of
universality.
The singular interaction $(\alpha > d-1)$
causes the
breakdown of both the Fermi and the Luttinger liquid pictures
where $|\Delta \nu_{\bf Q
\sigma}|=1$.
Instead, a new liquid is stable.
We call this liquid the {\it statistical spin liquid} because
the singular dynamical interaction is transmuted into the statistical
interaction between the particles with opposite spins and
 with the same ${\bf k}$
number \cite{s1,ja}.
This statistical interaction  removes half of the available single-particle
states from the
physical space and is characterized by
 the index of the Fermi surface singularity
$|\Delta \nu_{{\bf Q} \sigma}|=1/2$.
This liquid is an example of many-particle system with the fractional
exclusion statistics
\cite{hal}, which takes into account the spin degrees of freedom \cite{ja}.

In the low temperature regime one can perform the Sommerfeld expansion
for thermodynamic quantities.
For example,
the temperature dependence of the chemical potential in the paramagnetic case
has the form
\begin{equation}
\mu=\epsilon_F - k_BT\ln2 - \frac{\pi^2}{6}\left(\frac{\rho'}{\rho}\right)
(k_BT)^2 + O(T^4),
\end{equation}
where $\epsilon_F$ is the Fermi energy and $\rho$ ($\rho'$) is the value
of the density of states (its derivative)
taken at the Fermi energy.
In the statistical spin liquid case a linear in $T$ term appears.
This
unconventional type of $\mu$ dependence might be observed in
the thermopower measurements.
We have also calculated the internal energy per particle, which is
\begin{equation}
\frac{E}{N} = \frac{E_0}{N} + \frac{\pi^2}{6} (k_BT)^2 \rho + O(T^4),
\end{equation}
where $E_0$ is the ground state energy.
Thus, the
specific heat is
$C_V = \gamma _{sl} T$, where $\gamma_{sl} = (1/3)\pi^2k_B^2\rho$.
It is a linear function in $T$, as in the Fermi liquid
case, but the $\gamma$ coefficient is one half of that for the Fermi
liquid.
This is due to the  exclusion of the half of the total number of states.

It is also interesting to inquire how the system magnetization changes
for an arbitrary value of $U_s$.
Using the formula (7) we find that the magnetization of the system is
expressed by
$$
m=\frac{1}{N} \sum_{\bf k} \left[ \delta n_{\bf k \uparrow} -
\delta n_{\bf k \downarrow} \right] =
$$

\begin{equation}
\frac{1}{N} \sum_{\bf k}
\frac{ e^{\beta U_s} e^{\beta (\epsilon_{\bf k}-\mu)} \cosh (\beta h)}
{1+e^{\beta U_s}e^{\beta (\epsilon_{\bf k}-\mu)}
[e^{\beta (\epsilon_{\bf k}-\mu)}+2 \cosh(\beta h)]}
\tanh(\beta h).
\end{equation}
In Fig.3 we present our numerical solution for the magnetization per particle
as a function of the applied magnetic field  for selected values of
$U_s>0$.
The magnetization curve saturates faster for  larger $U_s$ values.
Also, in the statistical spin liquid case, only one-half of the magnetic
field is needed to saturate the moment, as compared to that for
the Fermi liquid.
This is easy to understand because in the present
case  each $\bf k$ state is singly occupied.
In the $U_s \rightarrow \infty$ limit the total magnetization
curve coincides with that for $nN$ localized moments \cite{s1,sok}.

The single particle Green function for the Fermi liquid has
simple poles which correspond to the energy of quasiparticles.
A marginal interaction renormalizes the spectral weight, as well as
the energy spectrum of quasiparticles, but the analytical structure of the
Green function  remains unaltered.
The Luttinger liquid Green function has the
branch cuts instead of simple poles\cite{mattis}.
Therefore, the quasiparticle spectrum is not present and only
the collective excitations emerge.
In the statistical spin liquid case the propagator has again
simple poles.
However,
the spectral weight function is renormalized even if the
marginal interactions are
not included.
Namely, one can easily find that the Green function for
the spin liquid
is of the form
\begin{equation}
G^{R(A)}_{\sigma}({\bf k},\omega)=\frac{1}{2 \pi}
\frac{1-n_{\bf k \bar{\sigma}}}{\omega -\epsilon_{\bf k} \pm i \delta}.
\end{equation}
To derive this propagator one utilizes the commutation algebra
of creation and annihilation operators
($b_{\bf k \sigma}^+$ and $b_{\bf k \sigma}$, respectively)
in which the singly occupancy
principle is incorporated explicitly; this yields
\begin{equation}
\{b_{\bf k \sigma}, b^{+}_{{\bf k}' \sigma'}\} = \delta_{{\bf k},{\bf k}'}
\left[
(1-n_{\bf k \bar{\sigma}})\delta_{\sigma, \sigma'} +
b_{{\bf k} \sigma}^{+}b_{{\bf k} \bar{\sigma}}(1-\delta_{\sigma, \sigma'})
\right].
\end{equation}
Since these operators do not commute to a number the statistical
factor appears in the numerator of the Green function.
Surprisingly enough, the interaction of the power $\alpha$ stronger then $d-1$
restores the quasiparticle structure again.
As in the Fermi liquid, we can include the marginal interaction
as a perturbation on the statistical spin liquid state.

One may view the singularity of the scattering amplitude from a
different prospective.
The double-occupancy projection may be regarded as a bone fide instability
of the Fermi surface (cf.Fig.1).
This is  clear to see in the case
with one particle per one available ${\bf k}$ state
(i.e. the case with one particle per atom).
Then, the ground state of the spin liquid is that of a Mott insulator,
since all available states are singly occupied and the configurational
entropy ($k_B\ln 2$ per particle) is that of a Mott insulator in
the spin disordered phase.
Therefore, the Fermi liquid - spin liquid boundary
may be considered as a Mott-Hubbard boundary for this particular band filling
$n=1$.
The scaling properties near this boundary as a function of $n$ have
been  studied recently \cite{cont} from the spin liquid side;
the corresponding situation from the Fermi liquid side has been studied
for the Hubbard model system both at $T=0$ long time ago \cite{br}
and for $T>0$ later \cite{s2}.
The Mott localization appears  as
an instability of the Fermi surface  against normal fluctuations
and is characterized by the condition
$k_{F{\bf Q}\sigma}^{0} \approx \delta k_{F {\bf Q} \sigma}({\bf x})$.
The nature of the collective excitations near this instability,
as well as the role of
the Luttinger liquid stability in the regime
 in between the Fermi- and the spin-liquid
states, all treated as separate condensed phases, needs to be explored.

In summary, we noted that the effective interaction with strong enough
singularity can lead to an instability of the Luttinger liquid  state
with respect to the statistical spin liquid state in the same manner as the
Fermi liquid becomes unstable with respect to the Luttinger liquid.
The three states: the Fermi, the Luttinger, and the  statistical
spin liquids are the only three universality classes arising from the
power-law scaling
behavior of the interaction between the quasiparticles.
We investigated also the statistical properties and the quasiparticle
spectrum of such spin liquid and compared them
with those for the Fermi and the Luttinger liquids.
We did not address, how such a  singular interaction
might arise.
Hatsugai and Kohmoto \cite{hat}
pointed to the long-range nature of the interaction as a
possible source of such a singular behavior.
The presence of the
van-Hove singularities on the Fermi surface  enhances the
scattering amplitude via the vanishing Fermi
velocities in Eq.(2), and may stabilize either the Luttinger or
the statistical spin liquids.

\vspace{1cm}

The work was  supported
by the Committee of Scientific Research (KBN) of Poland,
Grants Nos.  2 P302 171 06 and 2 P302 093 05.
The authors are also grateful to the Midwest Superconductivity Consortium
(MISCON) of U.S.A. for the support through Grant No. DE-FG 02-90 ER 45427.

\figure{Fig.1
Temperature dependence of the chemical potential for different values
of $U_s$. The limit $U_s\rightarrow \infty $ represents the
statistical spin liquid.}

\figure{Fig.2
Temperature dependence of the entropy. The $T=0$ value for $U_s/W >1$
is $k_B\ln2$ per particle.}

\figure{Fig.3 Magnetization curves for different $U_s$. The $U_s=\infty$ curve
corresponds to that for localized moments.}

%\end{narrowtext}
\end{document}